\documentclass[manuscript]{acmart}
\AtBeginDocument{%
  }

\setcopyright{acmlicensed}
\copyrightyear{2025}
\acmYear{2025}
\acmDOI{XXXXXXX.XXXXXXX}
\acmConference[Conference acronym 'XX]{Make sure to enter the correct
  conference title from your rights confirmation email}{June 03--05,
  2025}{Woodstock, NY}
\acmISBN{978-1-4503-XXXX-X/2025/06}
\usepackage[normalem]{ulem}
\usepackage{enumitem}
\usepackage{multicol}
\usepackage{multirow}
\usepackage{booktabs}
\usepackage{pifont}
\newcommand{\cmark}{\ding{51}}
\newcommand{\xmark}{\ding{55}}

\begin{document}

\title{Deep Research: A Survey of Autonomous Research Agents}


% \author{Wenlin Zhang$^1$, Xiaopeng Li$^{1}$, Yingyi Zhang$^{1,3}$, Pengyue Jia$^{1}$, Yichao Wang$^{2}$, Huifeng Guo$^2$, Xiangyu Zhao$^1$}
% \affiliation{
% 	\institution{$^1$City University of Hong Kong, $^2$Huawei Noah’s Ark Lab, \\ $^3$Dalian University of Technology}
% 	\country{}
% }
% \email{{wl.z,xiaopli2-c,x,jia.pengyue}@my.cityu.edu.hk, yingyizhang@mail.dlut.edu.cn, xianzhao@cityu.edu.hk}
% \email{{wangyichao5}@huawei.com}
% \thanks{* Equal contribution.}

\author{Wenlin Zhang}
\email{wl.z@my.cityu.edu.hk}
% \orcid{Your ORCID if available} % Uncomment and fill if you have one
\affiliation{%
  \institution{City University of Hong Kong}
  \country{China}
}

\author{Xiaopeng Li}
\email{xiaopli2-c@my.cityu.edu.hk}
% \orcid{Your ORCID if available}
\affiliation{%
  \institution{City University of Hong Kong}
  \country{China}
}

\author{Yingyi Zhang}
\email{yingyizhang@mail.dlut.edu.cn}
% \orcid{Your ORCID if available}
\affiliation{%
  \institution{Dalian University of Technology \& City University of Hong Kong}
  \country{China}
}

\author{Pengyue Jia}
\email{jia.pengyue@my.cityu.edu.hk}
% \orcid{Your ORCID if available}
\affiliation{%
  \institution{City University of Hong Kong}
  \country{China}
}

\author{Yichao Wang}
\email{wangyichao5@huawei.com}
% \orcid{Your ORCID if available}
\affiliation{%
  \institution{Huawei Noah’s Ark Lab}
  \country{China}
}

\author{Huifeng Guo}
\email{huifeng.guo@huawei.com}
% \orcid{Your ORCID if available}
\affiliation{%
  \institution{Huawei Noah’s Ark Lab}
  \country{China}
}

\author{Yong Liu}
\email{liu.yong6@huawei.com}
% \orcid{Your ORCID if available}
\affiliation{%
  \institution{Huawei Noah’s Ark Lab}
  \country{China}
}

\author{Xiangyu Zhao$^{\dagger}$}
\email{xianzhao@cityu.edu.hk}
% \orcid{Your ORCID if available}
\affiliation{%
  \institution{City University of Hong Kong}
  \country{China}
}
\thanks{$^{\dagger}$Corresponding Author.}


\renewcommand{\shortauthors}{Wenlin Zhang et al.}

\begin{abstract}
    % Large Language Models (LLMs) have demonstrated strong generalization and reasoning abilities, yet remain bounded by static, pretraining-era knowledge. Retrieval-Augmented Generation (RAG) provides a mechanism to extend LLMs with dynamic external information, but traditional RAG systems often suffer from a misalignment between retrieved content and generation needs. Recent advances in agentic RAG reposition the LLM as an autonomous agent capable of iteratively controlling its own retrieval and synthesis pipeline. Building on this foundation, two emerging paradigms—\textit{Deep Search} and \textit{Deep Research}—highlight the need for multi-step planning, interactive exploration, and structured content generation. This survey presents a comprehensive review of the Deep Research Agent landscape by decomposing it into four core components: \textit{Planning}, \textit{Question Developing}, \textit{Web Exploration}, and \textit{Report Generation}. We analyze key system architectures, benchmarks, and limitations, offering insights into the development of fully autonomous, reliable, and effective research-capable agents.
    The rapid advancement of large language models (LLMs) has driven the development of agentic systems capable of autonomously performing complex tasks. Despite their impressive capabilities, LLMs remain constrained by their internal knowledge boundaries.
    To overcome these limitations, the paradigm of \emph{deep research} has been proposed, wherein agents actively engage in planning, retrieval, and synthesis to generate comprehensive and faithful analytical reports grounded in web-based evidence. In this survey, we provide a systematic overview of the deep research pipeline, which comprises four core stages: planning, question developing, web exploration, and report generation. For each stage, we analyze the key technical challenges and categorize representative methods developed to address them. Furthermore, we summarize recent advances in optimization techniques and benchmarks tailored for deep research. Finally, we discuss open challenges and promising research directions, aiming to chart a roadmap toward building more capable and trustworthy deep research agents.

% We further organize existing approaches according to evolving capability requirements and agentic decision points, highlighting recent advances in reinforcement learning, tool-augmented planning, browser-based agents, and factuality-aware generation. 

\end{abstract}



% \ccsdesc[500]{Computing methodologies~Planning and scheduling}
% \ccsdesc[500]{Computer systems organization~Embedded and cyber-physical systems}
% \ccsdesc[500]{Human-centered computing~Collaborative and social computing}



\maketitle

\section{Introduction}

\begin{figure*}
    \centering
    \includegraphics[width=\linewidth]{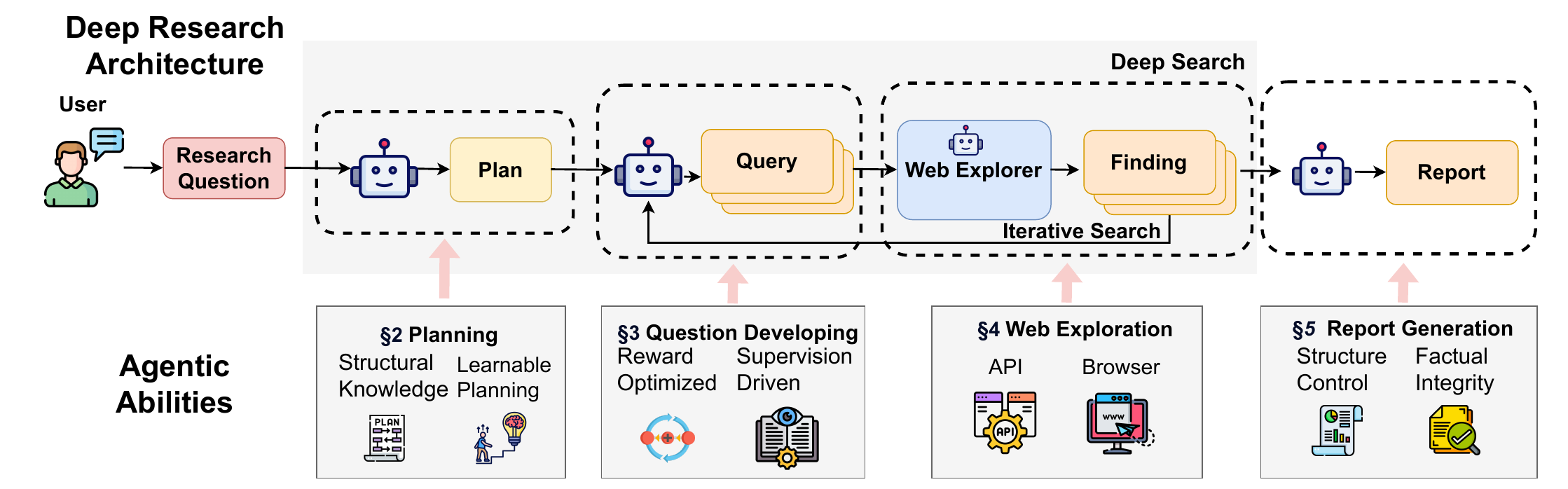}
    \caption{Overview of the deep research system.}
    \label{fig:overview}
\end{figure*}

Large Language Models (LLMs), such as GPT-4~\cite{openai2023gpt4}, Qwen3~\cite{yang2025qwen3}, and DeepSeek-R1~\cite{guo2025deepseek}, have achieved remarkable progress. These advances have enabled LLMs to serve as general-purpose language agents. However, LLMs remain constrained by their internal knowledge boundary, thereby limiting their effectiveness in dynamic or specialized scenarios.

To overcome these limitations, Retrieval-Augmented Generation (RAG) is proposed to augment LLMs reasoning with access to external information sources. RAG systems retrieve relevant documents from a large corpus and condition the LLM’s generation on the retrieved content. Although the RAG framework enhances factuality and adaptability, it constrains the model to passively consume retrieved content without participating in deeper exploration or reasoning. As LLMs continue to gain agentic abilities such as planning, tool use, and reflective reasoning, there is a growing shift from passive retrieval to active, goal-driven interaction with external knowledge sources. This shift marks the rise of a new paradigm known as \textit{agentic search}.

Within the agentic search paradigm, a prominent emerging approach is \textit{deep search}, where agents dynamically engage in planning, question developing, and web exploration to iteratively retrieve information aligned with evolving task objectives. While deep search significantly enhances the agent’s ability to acquire and contextualize knowledge, it remains insufficient for completing complex research tasks that demand synthesis and reasoning. To address this gap, the \textit{deep research} paradigm has been introduced, combining exploration with structured analysis and iterative generation. This paradigm enables agents to transform retrieved evidence into coherent, factual, and goal-aligned outputs, fulfilling the demands of high-level research workflows. The Deep Research process typically involves four interconnected stages:

\noindent \textbf{Planning.} This stage involves decomposing a high-level research question into structured sub-goals or subtasks. The agent must decide what to search for, in what order, and how intermediate information will support downstream synthesis. Unlike traditional step-by-step prompting, planning requires generating an explicit, task-aware roadmap before any retrieval or generation begins.

\noindent \textbf{Question Developing.} Given a sub-goal, the agent formulates one or more retrieval queries that capture specific, contextualized information needs. These queries may vary in abstraction, specificity, and granularity. In contrast to static question formulation~\cite{izacard2023atlas}, this stage requires adaptive generation of diverse and complementary queries tailored to evolving subgoals.

\noindent \textbf{Web Exploration.} The agent actively interacts with external sources either through web APIs or browser-based actions to collect relevant information. This process requires selecting tools, issuing queries, parsing results, and filtering noisy or redundant content. Unlike single-shot retrieval in traditional RAG pipelines, web exploration is iterative and agent-driven, enabling deeper coverage of sparse or scattered evidence.

% \noindent \textbf{Report Generation.} The retrieved information is integrated into a structural report. The agent must select supporting evidence, determine structure, and ensure consistency and completeness. Unlike extractive or shallow generation approaches \yc{it seems that this sentence separates the previous and the following sentences}, this step demands multi-source fusion, discourse planning, and task-specific formatting.

\noindent \textbf{Report Generation.} To produce a structured report, the agent must integrate retrieved information by selecting relevant evidence, organizing content coherently, and ensuring both consistency and completeness. This process goes beyond extractive or shallow generation, requiring multi-source fusion, discourse-level planning, and task-specific formatting.

As illustrated in Figure~\ref{fig:overview}, deep research systems involve multi-stage tasks and extensive information processing, which impose progressively evolving demands on agentic capabilities.

% \sout{The growing agentic capabilities of LLMs lay the foundation for building deep research systems. However, as illustrated in Figure~\ref{fig:overview}, the deep research system involves multi-stage tasks and extensive information processing, which impose evolving agentic capability demands and new challenges.} \yc{the logic seems not reasonable} 
Therefore, to make the best use of Deep Research systems, we organize the major challenges and corresponding solutions according to the key capability requirements:

\begin{itemize}[left=0pt, labelindent=0pt, itemindent=0pt]
    \item \textbf{Planning}: \textit{Given a research question, how to construct an effective and interpretable research plan before execution?} Planning serves as the first and foundational stage in deep research, translating user intent into actionable steps such as sub-question development and evidence retrieval. The key challenge lies in decomposing broad or ambiguous questions into a structured plan, especially under open-ended research goals. Unlike reactive agents that respond immediately, research agents must reason about intermediate steps and execution paths. This requires planning mechanisms that are both flexible to different question types and interpretable for human inspection. We define this process as the \textit{Planning} module (see Section~\ref{sec:planning}), which produces a structured, goal-driven plan to guide subsequent reasoning and retrieval.

    % \item \textbf{Planning}: \textit{Given a research question, how to construct an effective and interpretable research plan before execution.} In deep research agents, planning serves as the core stage that bridges user intent and downstream actions such as question development, web exploration, etc. However, transforming a natural language query into a structured plan—especially under open-ended or complex research goals—remains challenging. Unlike reactive agents that respond immediately, research agents must first understand the question, infer intermediate steps, and determine how to answer it. This demands planning mechanisms that are both flexible and explainable. 
    % % Drawing from recent developments in reasoning trajectory synthesis, plan refinement, and model-based simulation \yc{what's the necessity of this sentence}, 
    % We define this process as the \textit{Planning} module (see Section~\ref{sec:planning}), which produces a structured research plan conditioned on the question.
    \item \textbf{Question Developing}: \textit{Given a sub-goal from the planning stage, how to formulate effective, diverse queries that balance specificity and coverage.} In deep research agents, question developing is crucial for guiding the retrieval process to gather relevant information. However, generating queries that are both precise and comprehensive is challenging due to the need to capture specific information needs while ensuring broad coverage of the topic. Solutions typically involve adaptive query generation strategies, where agents dynamically formulate queries based on the evolving context of the research task. Techniques such as reinforcement learning are employed to optimize query effectiveness through interaction with the search environment, while non-RL methods like supervised fine-tuning or heuristic-based approaches provide alternative means to enhance query diversity and relevance. We define this process as the \textit{Question Developing} module (see Section~\ref{sec:question_developing}), which produces a set of targeted, contextualized queries conditioned on the sub-goals from the planning stage.
    % How to formulate effective, diverse queries that balance specificity and coverage.
    
    \item \textbf{Web Exploration}: \textit{Given a search plan and query, how to efficiently retrieve the most relevant and trustworthy information while filtering noisy, redundant, or conflicting information on the web?} In deep research workflows, efficiently locating and retrieving precise information online is critical for producing high-quality research reports. Recent works include autonomous web-retrieval agents that crawl hyperlinks to collect pertinent content~\cite{he2024webvoyager,yang2023mm,zhou2023webarena,nakano2021webgpt} and API-based methods that query search engines directly~\cite{bing,google,brave2022search,bochaai_open_platform}. 
    % \yc{add citations, Do we need such kind of description? existing solutions ..., more like a research paper, not a survey paper}. 
    We refer to this process as the \textit{Web Exploration} module (see Section~\ref{sec:web_retrieval}), which retrieves relevant information from diverse sources conditioned on the input research queries.

    % How to search efficiently while filtering noisy, redundant, or conflicting information.
    \item \textbf{Report Generation}: \textit{Given retrieved evidence related to the target research question, how to generate a coherent and structured report that keep factual integrity.} This stage is critical for transforming fragmented knowledge into a comprehensive analysis report. However, report generation faces challenges in structure control and factual integrity. To address these challenges, recent approaches incorporate structure-aware planning and constraint-guided generation to enforce layout coherence, and adopt evidence-grounded modeling to ensure factual reliability. We define this process as the \textit{Report Generation} module (see Section~\ref{sec:report_generation}), which produces structured, trustworthy report based on the information collected by deep research agents in web exploration stages.
\end{itemize}
% \yc{refine the aforementioned four paragraphs with coherent descriptions including goals, challenges}

In response to these challenges, we present a capability-centric survey of deep research agents, with a particular focus on how core capabilities—such as planning, retrieval, reasoning, and report generation—can be independently optimized and jointly integrated. Rather than enumerating full-system pipelines, we dissect the modular competencies underlying deep research and analyze their technical bottlenecks and coordination challenges. To this end, we first analyze key obstacles in building deep research agents; then we review representative methods across different modules, and finally highlight emerging trends, such as reasoning-driven retrieval, structured report generation, and self-evolving agents. Compared to prior surveys, our work offers a fine-grained, modular perspective. For example, Huang et al.\cite{huang2025deep} emphasize system architecture and future roadmaps, while Xu et al.\cite{xu2025comprehensive} provide a broad, enumerative overview of tasks and tools. In contrast, our taxonomy centers on capability formation and integration, aiming to reveal deeper connections between methods and provide insights into evolving design paradigms.

% In response to these challenges, we present a focused survey of the deep research system. We review representative methods, analyze core technical bottlenecks \yc{analyze xx first, review xx second?}, and \yc{provide something like insight xxx} highlight promising directions toward building more capable, autonomous, and knowledge-intensive LLM agents. As far as we know, this survey is fundamentally different from the existing two surveys. One prior survey~\cite{huang2025deep} focuses on system architecture and future roadmaps for deep research agents, while another~\cite{xu2025comprehensive} provides a broad overview of deep research systems and applications but leans toward enumerative listing rather than fine-grained comparison \yc{Is it necessary to mention other surveys in detail?}. In contrast, our survey offers a capability-centric, modular taxonomy of techniques, aiming to reveal the underlying connections among current methods and shed light on emerging trends \yc{missed in the following part?} in the field.

% \input{2Problem Definition}
\section{Planning}
\label{sec:planning}
Deep research systems increasingly emphasize a dedicated planning stage to orchestrate complex, long-horizon tasks. Rather than relying on reactive, step-by-step prompting alone, modern agents employ explicit planning to decide what actions or sub-tasks to execute before carrying them out. This planning process helps overcome the limitations of purely reactive prompting by improving task decomposition, guiding retrieval and reasoning more systematically, and reducing failures caused by trial-and-error strategies.

% \section{Planning}
% \label{sec:planning}
% Deep research systems increasingly emphasize a dedicated planning stage to orchestrate complex, long-horizon tasks. Rather than relying on reactive, step-by-step prompting alone, modern agents employ explicit planning to decide what actions or sub-tasks to execute before carrying them out. This approach addresses key challenges encountered by LLM agents in open-ended environments – from the web to code and even question generation – where naive trial-and-error or greedy strategies often fail to yield robust results.\yc{not concise and formal enough.}

\begin{definition}[Planning]
In deep research agents, \textit{planning} refers to the process by which an agent transforms a user query and prior knowledge into a structured research plan consisting of intermediate subgoals or actions. 
Formally, given an initial research question \( q_0 \) and agent context \( \mathcal{K} \), the planning model \( \mathcal{M}^{\mathrm{plan}}\) parameterized by $\theta$ produces a plan:
\begin{equation}
    \mathcal{P} = \mathcal{M}^{\mathrm{plan}}(q_0, \mathcal{K}; \theta),
\end{equation}
where \( \mathcal{P} = [s_1, s_2, \dots, s_n] \) denotes a sequence of subgoals or tool-invocation steps guiding downstream execution.

% \sout{
% In the context of deep research agents, \textit{planning} refers to the process by which an agent transforms a user query—along with its internal knowledge and parameters \yc{what is the difference between internal knowledge and parameters} of the language model—into a structured research plan composed of intermediate steps, subgoals, or tool-augmented actions.
% Formally, given the initial research question $q_{0}$ and agent memory $\mathcal{K}$ \yc{never mentioned before?}, the agent performs:
% \begin{equation}
%     \text{P} = \mathcal{M}_\theta(q_{0}, \mathcal{K}),
% \end{equation}
% where $\mathrm{LLM}(\cdot)$ \yc{should be $\mathrm{M}(\cdot)$?} denotes a language model $\mathcal{M}$ with parameters $\theta$ that are targeted for producing a sequence of reasoning steps, tool calls, or retrieval strategies that guide task execution.
% }
\end{definition}

This planning strategies may include world model simulation~\cite{gusimulate}, modular design search~\cite{shang2024agentsquare}, human-like reasoning trajectory synthesis~\cite{sun2025simpledeepsearcher}, or agent self-refinement~\cite{patel2024large}. To better understand the diversity of planning strategies, we summarize representative methods in Table~\ref{tab:planning}, categorized by their planning strategies. The plan is dynamically generated and may adapt to task complexity, resource availability, or user intent. It acts as the bridge between high-level goals and low-level execution in research-oriented web agents. 

\begin{table}[h]
\centering
\caption{Taxonomy of planning methods in deep research agents.}
\label{tab:planning}
\small
\begin{tabular}{l|l|p{7.5cm}}
\toprule
\textbf{Planning} & \textbf{Category} & \textbf{Related Works} \\
\midrule
\midrule
\multirow{3}{2.5cm}{Planning with Structured World Knowledge} 
    & Planning via simulation & WebDreamer~\cite{gu2024your}, Simulate Before Act~\cite{gusimulate} \\
% \cmidrule{2-3}
    & Planning via modularity & Webpilot~\citet{zhang2025webpilot}, WKM~\cite{qiao2024agent},Plan-and-Act~\citet{erdogan2025plan} \\
% \cmidrule{2-3}
    & Planning via adaptation &  Thought of Search~\cite{katz2024thought}, MPO~\cite{xiong2025mpo} \\
\midrule
\multirow{3}{2.5cm}{Planning as a Learnable Process} 
    & Planning via space exploration & Agentsquare~\cite{shang2024agentsquare}, Agent-E~\citet{abuelsaad2024agent} \\
% \cmidrule{2-3}
    & Planning via self-training & \citet{patel2024large}, InSTA~\citet{trabucco2025towards} \\
% \cmidrule{2-3}
    & Planning via preference modeling & MindSearch~\citet{chen2024mindsearch}, SimpleDeepSearcher~\cite{sun2025simpledeepsearcher}, Search-in-the-Chain~\cite{xu2024search}, WEPO~\cite{liu2025wepo}, MPO~\cite{xiong2025mpo} \\
\bottomrule
\end{tabular}
\end{table}

\subsection{Planning with Structured World Knowledge}
Research on web-based agents increasingly highlights the importance of leveraging \emph{structured world knowledge}—ranging from learned latent world models to explicit graphs—to guide long-horizon decision making. Early evidence suggests that large language models (LLMs) can act as \emph{implicit} world models, providing sufficient environment priors for goal-directed reasoning~\cite{gu2024your}. Building on this insight, the \textsc{Simulate Before Act} framework introduces an explicit simulation phase that enables agents to mentally roll out candidate action trajectories and evaluate their feasibility prior to execution, thereby improving robustness and adaptability~\cite{gusimulate}.  

Structured, multi-module pipelines further enrich this paradigm. WebPilot partitions web exploration across specialized sub-agents coordinated by a strategic planner, demonstrating how distributed planning built atop a shared world representation boosts task efficiency~\cite{zhang2025webpilot}. Qiao \emph{et al.} incorporate external knowledge graphs to ground reasoning paths, exemplifying \emph{knowledge-grounded} planning that explicitly models environment structure~\cite{qiao2024agent}. For temporally extended tasks, Erdogan \emph{et al.} couple an initial plan with iterative refinement steps to maintain coherence over long horizons~\cite{erdogan2025plan}.  

Efficiency considerations also shape contemporary methods: Katz \emph{et al.} quantify the computational cost of reasoning operations and propose an efficiency-aware planning strategy that balances deliberation depth against resource usage~\cite{katz2024thought}. Finally, Xiong \emph{et al.} present \textit{meta-plan optimization} (MPO), a meta-learning approach that adaptively tunes planning strategies across diverse web environments, closing the loop between world-model fidelity and strategic flexibility~\cite{xiong2025mpo}. Collectively, these works operationalize structured knowledge as a foundation for foresightful and computationally prudent web-agent planning.

\subsection{Planning as a Learnable Process}
A complementary research trajectory treats the \emph{planning mechanism itself} as the primary object of learning, allowing agents to iteratively refine their decision procedures through search, feedback, and large-scale training. AgentSquare exemplifies this view by performing architecture search over a vast configuration space to automatically assemble task-adapted planning pipelines, dispensing with hand-crafted heuristics~\cite{shang2024agentsquare}. Agent-E distills reusable components from navigation tasks, showing how robust behavior can emerge from system-level decomposition~\cite{abuelsaad2024agent}.  

Self-improvement via interaction feedback has also proven effective: Patel \emph{et al.} demonstrate that LLM-based agents can hone their planning heuristics purely from deployment experience, without human intervention~\cite{patel2024large}. Scaling this idea, Trabucco \emph{et al.} propose internet-scale training pipelines that generalize planning behaviours across a massive distribution of tasks, shifting attention from per-task optimization toward lifelong adaptability~\cite{trabucco2025towards}.  

Reasoning-enhanced planning mechanisms further enrich this learnable paradigm. MindSearch simulates human-like search strategies to deepen planning trajectories~\cite{chen2024mindsearch}, whereas SimpleDeepSearcher synthesizes multi-hop reasoning paths for information-seeking tasks~\cite{sun2025simpledeepsearcher}. Search-in-the-Chain closes the loop between retrieval and planning by dynamically updating action choices based on intermediate search results~\cite{xu2024search}. WEPO tailors planning to user interface preferences, optimizing action sequences at the UI-element level~\cite{liu2025wepo}. MPO, originally introduced for meta-learning across environments, also exemplifies how agents can refine planning strategies via higher-order gradient updates~\cite{xiong2025mpo}.  

Together, these studies reconceptualize planning as a \emph{learnable, evolvable capability}, foregrounding continual self-improvement and task-driven adaptation over static execution.

\subsection{Discussion}
The planning stage has become a defining component in recent deep research agents, offering not only structure to downstream reasoning but also interpretability and control in open-ended tasks. One notable strength is the emergence of explicit, structured planning outputs—such as reasoning chains, search commands, or subgoal sketches—which help agents avoid short-sighted or repetitive behaviors.  Another encouraging trend is the agent’s ability to learn and refine planning strategies over time. Through interaction feedback, meta-learning, or exposure to large-scale task distributions, agents are beginning to self-adjust their planning routines, yielding increasingly robust behaviors in dynamic web environments. 

However, important limitations remain. First, plans generated by current LLMs are often brittle, lacking robustness to ambiguous research questions or underspecified goals. Even with structured formats, the internal consistency of plans is not guaranteed, and hallucinated steps can propagate errors downstream. Second, while planning modules are increasingly modular and learnable, their evaluation remains coarse, often relying on end-task accuracy rather than plan quality itself—making it hard to diagnose planning failures or compare strategies meaningfully. Third, many systems treat each research question as an isolated problem, without leveraging shared structures or transferable strategies, which limits the agent’s ability to accumulate generalizable planning knowledge across tasks. 
% \yc{is there work to resolve these challenges or some advices?}

Indeed, there is growing research aimed at addressing these challenges—from the introduction of standardized benchmarks like DeepResearch Bench \cite{du2025deepresearch} for evaluating report fidelity and citation accuracy, to reinforcement learning frameworks such as DeepResearcher \cite{zheng2025deepresearcher} that promote iterative self‑reflection and improved planning in web environments. But key limitations remain, particularly in handling ambiguity, ensuring long-range consistency, and transferring planning strategies across tasks.
\section{Question Developing} ~\label{sec:question_developing}

To support deep research tasks such as multi-step reasoning and information synthesis, it is often insufficient to rely on a single static query. Instead, systems must dynamically generate more targeted, contextualized, or decomposed queries that elicit useful evidence from retrieval modules or the web. This process, known as \emph{question developing}, is essential for guiding agentic retrieval strategies that operate beyond keyword matching and towards task-aware information seeking. 
% Based on how the query is reformulated and participates in retrieval, existing methods can be broadly categorized into two types: explicit query reformulation, where the original question is directly rewritten into a new query that is then used for retrieval; and dynamic in-context query reformulation, where the model dynamically generates queries as needed during the reasoning process, inserting them into the chain of thought such that querying and reasoning are interleaved.

\subsection{Definition}

% Given an initial query $q_0$, a task context $T$, and a search environment $\mathcal{E}$, \textbf{question developing} is the process of generating a set of transformed queries $\mathbf{Q} = \{q_1, q_2, \dots, q_k\}$ by selecting a query development policy $\pi$ from a policy space $\Pi$, such that the expected utility of retrieved information is maximized:
% \begin{equation}
%     \pi^* = \arg\max_{\pi \in \Pi} \; \mathbb{E}_{\mathbf{Q} \sim \pi(q_0, T)} \left[ \mathcal{U}(\mathcal{R}(\mathbf{Q}; \mathcal{E}), T) \right],
% \end{equation}
% where $\pi$ is a query developing policy that maps $(q_0, T)$ to one or more reformulated queries, $\mathcal{R}(\mathbf{Q}; \mathcal{E})$ is the retrieval operator over the environment $\mathcal{E}$ using $\mathbf{Q}$, $\mathcal{U}(\cdot, T)$ is a task-specific utility function measuring how useful the retrieved evidence is for completing the research task $T$.

\begin{definition}[Question Developing]
\textit{Question developing} denotes the process of converting each subgoal \( s_i \) within a structured plan \( \mathcal{P} = [s_1, \dots, s_n] \) into a series of search queries \( \mathcal{Q}_i = \{q_{i,1}, q_{i,2}, \dots \} \). This process is guided by the overall plan \( \mathcal{P} \), the current subgoal \( s_i \), and the accumulated evidence \( \mathcal{E} \), which comprises information retrieved from previous queries. Formally, given a query generation model \( \mathcal{M}^{\mathrm{ask}} \) with parameters \( \theta \), the queries for \( s_i \) are generated as
\begin{equation}
    \mathcal{Q}_i = \mathcal{M}^{\mathrm{ask}}(\mathcal{P}, s_i, \mathcal{E}; \theta).
\end{equation}

The retrieval process that produces and updates the evidence set \( \mathcal{E} \) is detailed in Section~\ref{sec:web_retrieval}.
\end{definition}

% \sout{In deep research agents, \textbf{question developing} refers to the process of mapping a multi-step plan $P = [s_1, s_2, \dots, s_n]$ into a set of executable search queries $\mathcal{Q} = \{q_1, q_2, \dots, q_n\}$, where each $q_i$ is dynamically generated by a language model conditioned on the subgoal $s_i$ \yc{seems not appear in the formulation, or related to $P$} and previously retrieved evidence. Formally, the query set is given by:
% \begin{equation}
%     \mathcal{Q} = \mathcal{M}_\theta(P, \mathcal{R}, \mathcal{E}, s_i),
% \end{equation}
% where $\mathcal{M}_\theta$ denotes the LLM with parameter $\theta$, which recursively consumes the plan $P$, interacts with a retrieval operator $\mathcal{R}$ over environment $\mathcal{E}$, and produces a sequence of queries $\mathcal{Q}$.}
% This process supports stepwise grounding, evidence-aware reformulation, and plan-conditioned query synthesis, serving as the critical link between high-level planning and low-level information acquisition.

% \yc{rewrite, categorization should also be reasonable or explainable, e.g., the effectiveness or efficiency can be impacted by the training paradigm, more details} 
Question developing methods can be categorized into two types based on optimization methods: Reward-Optimized Methods, where the query formulation process is optimized through trial-and-error exploration guided by a reward signal; and Supervision-Driven Methods, which rely on supervised finetuning or manually designed strategies without explicit reward-driven query optimization. A taxonomy of these approaches is provided in Table~\ref{tab:question-developing}, organized by training paradigm and key modeling strategies. In the remainder of this section, we formalize the notion of question developing and examine representative strategies that instantiate this process.

\begin{table}[h]
\centering
\small
\caption{Taxonomy of question developing methods in deep research agents.}
\label{tab:question-developing}
\begin{tabular}{l|l|p{7cm}}
\toprule
\textbf{Optimization} & \textbf{Category} & \textbf{Related Works} \\
\midrule
\midrule
% \multirow{4}{*}{RL-based Methods} 
\multirow{4}{*}{Reward-Optimized Methods}
    & Rewards for format and accuracy & 
    \textsc{DeepResearcher}~\cite{zheng2025deepresearcher}, \textsc{EvolveSearch}~\cite{zhang2025evolvesearch}, \textsc{R1-Searcher}~\cite{song2025r1}, \textsc{Search-R1}~\cite{jin2025search}, \textsc{ZeroSearch}~\cite{sun2025zerosearch}, \textsc{MaskSearch}~\cite{wu2025masksearch}, \textsc{DeepRetrieval}~\cite{jiang2025deepretrieval} \\
% \cmidrule{2-3}
    & Multi-dimensional reward & 
    \textsc{InForage}~\cite{qian2025scent}, \textsc{OTC-PO}~\cite{wang2025acting}, \textsc{IKEA}~\cite{huang2025reinforced}, \textsc{AutoRefine}~\cite{shi2025search}, \textsc{R-Search}~\cite{zhao2025r}, \textsc{MMSearch-R1}~\cite{wu2025mmsearchr1incentivizinglmmssearch}, \textsc{VRAG-RL}~\cite{wang2025vrag} \\
\midrule
% \multirow{2}{*}{Non-RL Methods} 
\multirow{2}{*}{Supervision-Driven Methods}
    & Multi-agent systems & 
    \textsc{ManuSearch}~\cite{huang2025manusearch}, \textsc{Search-o1}~\cite{li2025search}, \textsc{SearchAgent-X}~\cite{yang2025demystifying} \\
% \cmidrule{2-3}
    & Supervision optimization & 
    \textsc{ReasonRAG}~\cite{zhang2025process} \\
\bottomrule
\end{tabular}
\end{table}

\subsection{Reward-Optimized Methods}

Reinforcement learning (RL) provides a principled framework for optimizing query generation policies through interaction with a search environment and feedback from task-specific rewards. Unlike supervised fine-tuning, which requires labeled query-response pairs, reward-optimized approaches allow agents to explore query strategies and adaptively adjust when and how to search based on end-task outcomes such as answer accuracy, retrieval coverage, or efficiency. 

A common design choice in many RL-based question developing methods is to define the reward purely based on output format correctness and final answer accuracy. These methods treat retrieval and reasoning as black-box components, optimizing only for whether the output conforms to a specified format and leads to a correct answer. This simplifies the reward signal and makes it easier to scale training in noisy or simulated environments.

For instance, \textsc{DeepResearcher}\cite{zheng2025deepresearcher} and \textsc{EvolveSearch}\cite{zhang2025evolvesearch} use binary format rewards and token-level F1-based answer rewards, training multi-agent systems to interact with real or simulated search APIs via GRPO. \textsc{R1-Searcher}\cite{song2025r1} further introduces a staged training framework where the first phase rewards only format compliance and search usage, and the second phase integrates F1-based answer quality. Similarly, \textsc{Search-R1}\cite{jin2025search} enforces strict format templates (e.g., <think>, <search>, <information>, <answer>) and optimizes an exact-match-based reward via PPO or GRPO. \textsc{ZeroSearch}\cite{sun2025zerosearch} follows the same principle in a simulated setting by replacing real search APIs with a learned search simulator, thus enabling reward learning with curriculum rollout without any real querying cost. \textsc{DeepRetrieval}\cite{jiang2025deepretrieval} defines the reward over retrieval metrics such as Recall@K or NDCG@K, aligning query generation with retrieval performance. Lastly, \textsc{MaskSearch}\cite{wu2025masksearch} defines rewards purely based on masked span recovery (via retrieval-augmented answer prediction), combining format and accuracy into a unified reward function during pretraining. These reward designs help guide the agents toward generating queries that are not only syntactically valid but also more likely to retrieve answer-relevant content, especially when operating in noisy or weakly supervised environments. 
% \yc{seems that all of the work in this paragraph can also be placed in other sections, emphasizing the impact for query developing}

In contrast to approaches that define reward solely based on output format and answer correctness, a second class of question developing methods integrates richer, multi-dimensional reward signals that better reflect the complexities of interactive reasoning and search~\cite{song2025r1}. These reward functions not only consider what answer is produced, but also how it is obtained—penalizing unnecessary search, rewarding informative intermediate steps, or adapting to the agent’s internal vs. external knowledge boundaries.

\textsc{InForage}\cite{qian2025scent} augments outcome-based reward with \emph{information gain} (coverage of ground-truth knowledge) and an \emph{efficiency penalty} that discourages redundant reasoning hops. \textsc{OTC-PO}\cite{wang2025acting} introduces a cost-aware reward that penalizes excessive tool usage, aiming to minimize external calls while preserving correctness. \textsc{IKEA}~\cite{huang2025reinforced} further incorporates a \emph{knowledge boundary-aware reward}, giving agents positive feedback for solving “easy” questions using internal knowledge alone, and penalizing unnecessary or unproductive external searches. Other works emphasize reward shaping over multi-turn trajectories. \textsc{AutoRefine}\cite{shi2025search} evaluates not only final answer quality but also the completeness of intermediate refinements extracted from retrieved documents. \textsc{R-Search}\cite{zhao2025r} and \textsc{MMSearch-R1}\cite{wu2025mmsearchr1incentivizinglmmssearch} combine answer correctness with evidence quality and format compliance, supporting structured reasoning across modalities. Finally, \textsc{VRAG-RL}~\cite{wang2025vrag} adapts this idea to vision-language settings, optimizing visual search actions via fine-grained rewards for image selection, visual attention consistency, and final answer quality.

These reward-optimized methods demonstrate the viability of learning adaptive and interpretable query development policies in noisy, dynamic, and high-stakes research environments. They lay the groundwork for agents that can reason about the utility of queries in context and improve their behavior over time through reinforcement feedback.

\subsection{Supervision-Driven Methods}

In contrast to reward-optimized approaches, which rely on RL optimization, supervision-driven methods develop question generation strategies using supervised fine-tuning, rule-based decompositions, or structured multi-agent workflows. These methods avoid the complexity and instability of reinforcement learning, instead leveraging human demonstrations, task-specific heuristics, or architectural optimizations to guide the development of effective query strategies.

One line of work focuses on building structured multi-agent systems that divide the question developing process into modular roles. \textsc{ManuSearch}\cite{huang2025manusearch} implements a transparent, open-source framework with separate agents for planning subquestions, conducting external web search, and extracting structured evidence from HTML content. Each agent operates based on deterministic or supervised rules, coordinating over multiple turns to perform complex query decomposition and information synthesis. Similarly, \textsc{Search-o1}\cite{li2025search} and \textsc{SearchAgent-X}~\cite{yang2025demystifying} explore system-level enhancements, such as adaptive retrieval scheduling and efficient batch processing, to improve the throughput and responsiveness of retrieval-augmented reasoning without changing the underlying language model's behavior.

Other methods explore imitation learning or preference-based optimization to supervise the question developing process. \textsc{ReasonRAG}~\cite{zhang2025process} replaces sparse, outcome-only supervision with fine-grained feedback over intermediate steps. Using Monte Carlo Tree Search (MCTS) to explore reasoning trajectories and Direct Preference Optimization (DPO) to rank them, ReasonRAG identifies more effective planning behaviors without relying on trial-and-error reinforcement. This allows the model to learn how to interleave search and reasoning more efficiently from fewer training samples.

% These supervision-driven methods offer several practical advantages: they are typically easier to train, require fewer environmental interactions, and are often more interpretable due to their modularity or reliance on explicit supervision. However, their generalization ability may be constrained by the diversity and quality of available supervision signals. As such, they form a complementary space to reward-optimized approaches, particularly well-suited for settings where data efficiency, transparency, or system integration are prioritized.\yc{general description for deep research, not for query developing specifically.}

These supervision-driven methods offer several practical advantages for question developing: they enable more controllable query generation, often producing well-structured and semantically faithful queries by mimicking human-written examples or following rule-based templates. Since they avoid interacting with noisy or expensive search environments during training, they are easier to optimize and more stable in low-resource scenarios. Furthermore, their explicit supervision makes it easier to diagnose and adjust query behavior in response to planning errors or task-specific failures. However, their effectiveness is often limited by the quality and coverage of available demonstrations; without sufficient diversity, the generated queries may fail to adapt to unseen subgoals or novel reasoning contexts. Overall, these methods complement reward-optimized approaches by offering safer and more interpretable solutions in settings where data efficiency, query consistency, and development simplicity are prioritized.

\subsection{Discussion}
The Question Developing module is a core component in deep research agents, responsible for transforming subgoals into a series of specific retrieval queries. These queries need to accurately reflect the intent of the subgoals while being broad enough to retrieve comprehensive and relevant information from external sources. It essentially serves as the starting point for the agent's exploration of the information space, directly impacting the quality of subsequent retrieval and answer generation. Currently, in the field of Question Developing, the main trends include reward-optimized methods and supervision-driven methods. Reward-optimized methods dynamically adjust query strategies through interaction with the search environment, using feedback signals (such as the relevance of retrieval results) to improve efficiency and accuracy. On the other hand, supervision-driven methods, such as rule-based query generation or multi-agent collaboration systems, enhance query diversity and specificity through predefined logic or teamwork. These methods are all aimed at enabling agents to "ask" more intelligently to meet complex research needs.

Despite progress, existing methods still have significant shortcomings. First, the generated queries often rely too heavily on the clarity of the subgoals; once the subgoals are vague or ambiguous, the quality of the queries drops significantly, leading to deviations in information retrieval. Second, many systems lack contextual coherence when generating queries, such as failing to effectively integrate previous query history or task background, resulting in repetitive or redundant questions. Finally, current methods perform poorly in handling open-ended problems, tending to generate overly narrow queries that fail to capture a wide range of potential information, limiting the agent's exploratory capabilities.
\section{Web Exploration}
\label{sec:web_retrieval}

Deep-research workflows, ranging from scientific discovery and literature review to fact-checking and other expert investigations, critically depend on retrieving precise, context-aware, and trustworthy evidence from the vast, heterogeneous information available on the web. Yet because relevant material is sparsely distributed across countless web pages, designing agents that can accurately locate and extract the most useful content remains a critical challenge. 

% In the remainder of this section, we first formalize the notion of \emph{web exploration} and then examine these two solution classes in detail.

\begin{definition}[Web Exploration]
In the context of deep research, \emph{web exploration} is the process of retrieving the most relevant information from online sources. This may involve (i) deploying an agent that recursively follows hyperlinks and filters content, or (ii) invoking search engine APIs to obtain ranked results of relevant web documents. Given the query $\mathcal{Q}_i$, web retriever $\mathcal{R}$, the web agent $\mathcal{M^{\mathrm{web}}}$ parameterized by $\theta$, and the open-web corpus $\mathcal{H}$, the web exploration process is defined as

\begin{equation}
    \mathcal{D} \;=\; \mathcal{M^{\mathrm{web}}} \bigl(\mathcal{R}, \mathcal{Q}_i, \mathcal{H}; \theta\bigr),
\end{equation}

where $\mathcal{D}$ denotes the set of documents ultimately retrieved.
\end{definition}

% Contemporary web-retrieval methods fall into two broad categories: (1) Web retrieval agents \yc{agent-based ?}: autonomous agent systems that conduct web retrieval including operations like browse, click, and extract information much like a human researcher. (2) API-based retrieval systems\yc{what's the hell system? api-based?}: directly using the current web search engine (e.g., Google Search, Bing Search, etc.) that let developers pull ranked documents or snippets directly into research pipelines. A summary of representative methods in both categories is presented in Table~\ref{tab:web-exploration}.

Contemporary web-retrieval methods fall into two broad categories: (1) Web agent-based systems: autonomous agent systems that conduct web retrieval including operations like browse, click, and extract information much like a human researcher. (2) API-based retrieval systems: directly using the current web search engine (e.g., Google Search, Bing Search, etc.) that let developers pull ranked documents or snippets directly into research pipelines. A summary of representative methods in both categories is presented in Table~\ref{tab:web-exploration}.

\begin{table}[h]
\centering
\small
\caption{Taxonomy of web exploration methods in deep research agents.}
\label{tab:web-exploration}
\begin{tabular}{l|l|p{7cm}}
\toprule
\textbf{Information Source} & \textbf{Category} & \textbf{Related Works} \\
\midrule
\midrule
\multirow{3}{*}{Web-based} 
    & Web scraping and crawling & Scrapy~\cite{scrapy}, BeautifulSoup~\cite{beautifulsoup4} \\
% \cmidrule{2-3}
    & Browser-based web agents & WebGPT~\cite{nakano2021webgpt}, Selenium~\cite{selenium} \\
% \cmidrule{2-3}
    & Multimodal web agents & WebVoyager~\cite{he2024webvoyager}, MM-ReAct~\cite{yang2023mm}, WebArena~\cite{zhou2023webarena} \\
\midrule
\multirow{2}{*}{API-based} 
    & Industrial search engines & Bing~\cite{bing}, X posts~\cite{xposts}, Google~\cite{google} \\
% \cmidrule{2-3}
    & Domain-specific search engines & Reportify~\cite{reportify_nd}, YanXueZhiDe~\cite{cnki2024_xai}, CNKI~\cite{cnki_nd}, DuckSearch~\cite{DuckSearch}, BraveSearch~\cite{brave2022search}, Bocha~\cite{bochaai_open_platform} \\
\bottomrule
\end{tabular}
\end{table}

\subsection{Web retrieval agents}
Web retrieval agents are autonomous systems that extract information from the web, forming a critical component of deep research systems. These agents have evolved from simple extractors to sophisticated multimodal systems capable of navigating complex interactive content and synthesizing information across diverse sources.

\noindent\paragraph{Browser-Based Autonomous Web Agents.} AI-driven browser agents marked a fundamental shift in web-based information gathering for deep research. Unlike static scrapers, these agents dynamically navigate web interfaces through real or simulated browsers, making contextual decisions about which paths to explore based on encountered content. WebGPT~\cite{nakano2021webgpt} pioneered this approach using a text-based browser that converted HTML into structured representations, enabling the language model to issue high-level commands such as \texttt{Find} \emph{keyword} to retrieve relevant passages. More sophisticated implementations leverage real browsers via Selenium~\cite{selenium} to query the Document Object Model directly or execute custom JavaScript. Agent-E exemplifies the compact-representation approach, constructing accessibility trees that preserve semantic structure while removing extraneous elements. These agents perceive content through textual representations and execute actions programmatically—clicking links, completing forms, or triggering interactive elements. While effective for text-heavy research tasks, this approach fails when critical information appears in visual layouts or embedded visualizations, motivating the transition to multimodal agents.

\noindent\paragraph{Multimodal Web Agents.} Current web retrieval agents integrate visual perception to process both textual content and visual cues—interpreting charts, recognizing interface patterns, and understanding spatial information organization. This capability proves essential for deep research requiring analysis of data visualizations or navigation of visually complex research databases. Two approaches have emerged: specialized prompting frameworks like MM-ReAct~\cite{yang2023mm} and fully integrated vision-language agents. Both combine rendered screenshots with textual metadata for comprehensive page representation. WebVoyager~\cite{he2024webvoyager}, using GPT-4V, combines screenshot analysis with HTML processing to achieve 59\% task success on real-world benchmarks, significantly outperforming text-only baselines. MM-ReAct~\cite{yang2023mm} embeds screenshots directly into reasoning chains, enabling iterative action execution based on visual feedback. Sightseer~\cite{laurenccon2024unlocking} reconstructs HTML/CSS from screenshots, while WebArena~\cite{zhou2023webarena} provides comprehensive evaluation frameworks for diverse interface patterns.

These multimodal agents demonstrate essential research capabilities—answering complex queries with citations and synthesizing findings from heterogeneous sources. Despite challenges in robustness and reliability, progress points toward agents that navigate web knowledge with expert-like strategies, transforming how researchers access and synthesize information. As these systems mature, they shift human effort from mechanical information gathering to higher-level analysis and insight generation.
% \yc{focus on explaining web exploration, conditioned on the role in deep research, currently too general and lack of correlation with other modules.}

\subsection{API-Based Retrieval Systems}

API-based retrieval enables rapid incorporation of external knowledge into deep research pipelines. By exposing standardized endpoints, mature industrial search engines can be seamlessly integrated, yielding search results that are both reliable and trustworthy. OpenAI DeepResearch~\cite{openai2025deepresearch} leverages Microsoft Bing’s web-search infrastructure~\cite{bing} to issue queries, extract passages, spawn follow-up queries, rank candidate documents, and even execute sandboxed code. Grok DeepSearch~\cite{xai2025grok3} operates its own crawler and supplements public web data with privileged access to X posts~\cite{xposts}. Gemini DeepResearch~\cite{gemini2024deepresearch} invokes Google’s proprietary search stack~\cite{google}, whereas Perplexity DeepResearch~\cite{perplexity2025deepresearch} employs a hybrid solution that fuses a Bing-style web index with Perplexity’s Sonar API, which provides BM25, keyword, and dense-vector reranking. Several domain-specific systems further narrow the scope to specialized corpora, such as Reportify~\cite{reportify_nd} integrates licensed market-research reports and authoritative financial news, while YanXueZhiDe~\cite{cnki2024_xai} offers academic retrieval over the CNKI~\cite{cnki_nd} repository. Open-source frameworks, in contrast, typically depend on third-party engines such as DuckSearch~\cite{DuckSearch}, BraveSearch~\cite{brave2022search}, or Bocha~\cite{bochaai_open_platform}.

\subsection{Discussion}

Web exploration is vital for deep research, providing precise and relevant evidence from the expansive web through two main approaches: Web retrieval agents excel at dynamically navigating web interfaces, mimicking human browsing to access interactive or unindexed content, making them highly adaptable for complex inquiries. However, their resource-intensive nature and lack of real-time trustworthiness assessment pose significant drawbacks. Conversely, API-based systems offer rapid, efficient access to pre-indexed data from established search engines, ensuring reliability but often overlooking niche or dynamic content. Both approaches, while effective in isolation, are limited by their disconnect from upstream research stages like question formulation and evidence planning, hindering their ability to fully adapt to specific research needs.
  
% \sout{Looking forward, the future of web exploration lies in hybrid models that dynamically integrate strengths of web agents and API-based systems, combining rapid retrieval with deep, interactive analysis. Such integration could enhance efficiency and flexibility, addressing current limitations. Furthermore, advancements in multimodal processing and real-time verification are critical to keep pace with the web’s evolving landscape, ensuring consistency and reliability across diverse sources. By bridging these gaps, web exploration can evolve to meet the rigorous demands of deep research, delivering more precise and trustworthy evidence retrieval.
% \yc{some core challenges in web agent are not issued. e.g. evidence extraction, correctness verification, ..., many module can be categorized according to the challenges?}}

The future of web exploration lies in hybrid architectures that systematically address these core challenges through integrated solutions combining the strengths of both approaches. Advanced systems must incorporate specialized modules for evidence extraction, correctness verification, and content quality assessment, enabling rapid initial retrieval followed by deep interactive analysis. Critical developments in multimodal processing and real-time verification frameworks will prove essential for maintaining pace with the evolving web landscape. By establishing robust categorization frameworks for different system modules and systematically addressing these technical challenges, web exploration can mature into a comprehensive foundation for deep research, delivering more precise and trustworthy evidence retrieval capabilities.
\section{Report Generation}
\label{sec:report_generation}
In deep research, generation extends beyond traditional QA tasks, aiming to produce a comprehensive and analytical report. This process, known as \emph{report generation}, seeks to synthesize fragmented information retrieved from the web into a report that is coherent in structure, logically organized, and faithful to the underlying evidence.

\begin{definition}[Report Generation]
In the context of deep research agents, \textit{report generation} refers to the process by which an agent synthesizes a comprehensive report integrated by web information. 
Given the initial research question $q_0$, the research plan $\mathcal{P}$, the retrieved documents$\mathcal{D}$, and the web agent $\mathcal{M}_{\theta}$, the report generation process can be defined as 
\begin{equation}
    \mathcal{Y} = \mathcal{M}_{\theta} (q_{0}, \mathcal{P}, \mathcal{Q}, \mathcal{D})
\end{equation}

where $\mathcal{Y}$ denotes the generated report.
\end{definition}
The methods for report generation can be broadly categorized into two types: \emph{structure control} and \emph{factual integrity}. (1) Structure control focuses on organizing multi-step reasoning and retrieved content into coherent formats, often relying on planning-aware generation or constraint-guided generation. (2) Factual integrity aims to ensure that the generated report is faithful to the retrieved evidence, typically through grounding mechanisms or post-generation verification. Representative methods under each category are summarized in Table~\ref{tab:report-generation}.

\begin{table}[h]
\centering
\caption{Taxonomy of artifact generation methods in deep research agents.}
\label{tab:report-generation}
\small
\begin{tabular}{l|l|p{8cm}}
\toprule
\textbf{Field} & \textbf{Category} & \textbf{Representative Works} \\
\midrule
\midrule
\multirow{5}{*}{Structure Control} 
    & Planning-based Generation & Agent Laboratory~\cite{schmidgall2025agent}, AI Scientist v2~\cite{yamada2025ai}, LongEval~\cite{wu2025longeval}, LongWriter~\cite{bai2024unleashing}, LongDPO~\cite{ping2025longdpo} \\
% \cmidrule{2-3}
    & Constraint-guided Generation & WebThinker~\cite{li2025webthinker}, Suri~\cite{pham2024suri}, \citet{wan2025cognitive} \\
% \cmidrule{2-3}
    & Structural-aware Evaluation & Long2RAG~\cite{qi2024long2rag}, ExPerT~\cite{salemi2025expert}, \citet{long2025beyond}, \citet{kim2025say}, \citet{huang2024calibrating} \\
\midrule
\multirow{3}{*}{Factual Integrity} 
    & Faithful Modeling & RAGSynth~\cite{shen2025ragsynth}, BRIDGE~\cite{dai2025after}, \citet{zhou2023context}, \citet{shi2024trusting} \\
% \cmidrule{2-3}
    & Conflict Reasoning & FaithfulRAG~\cite{zhang2025faithfulrag}, DRAGged~\cite{cattan2025dragged}, \citet{yuan2024discerning}, \citet{ying2023intuitive} \\
% \cmidrule{2-3}
    & Factuality Evaluation & Face4RAG~\cite{xu2024face4rag}, SFR-RAG~\cite{nguyen2024sfr}, \citet{wallat2024correctness}, FaithJudge~\cite{tamber2025benchmarking}, RAG-QA Arena~\cite{han2024rag}, MT-RAIG~\cite{seo2025mt} \\
\bottomrule
\end{tabular}
\end{table}

\subsection{Structure Control}
Structure control refers to generating long-form outputs that are both structurally coherent and globally consistent. These long-form outputs often span multiple sections and require effective planning, topical alignment, and layout adherence. Recent research has tackled this problem from three perspectives: planning-based generation, constraint-guided generation, and structure-aware alignment.

Planning-based generation focuses on organizing content before or during generation by leveraging document outlines or hierarchical decomposition. Agent Laboratory~\cite{schmidgall2025agent} applies paragraph-level planning strategies with structure-aware prompting, while AI Scientist v2~\cite{yamada2025ai} performs recursive tree-structured planning to maintain global layout consistency. In parallel, LongEval~\cite{wu2025longeval} and LongWriter~\cite{bai2024unleashing} exemplify this approach by decomposing the generation process into high-level outline planning and section-level synthesis. This hierarchical strategy helps maintain coherence across thousands of tokens. LongDPO~\cite{ping2025longdpo} further incorporates critique-augmented supervision at each generation step, ensuring local completions are aligned with the global document structure. 

Constraint-guided generation aims to enforce specific format, style, or content requirements during decoding. WebThinker~\cite{li2025webthinker} introduces section-aware decomposition that bridges the planning and generation stages, enabling the model to map structured subtasks to content sections. Suri~\cite{pham2024suri} investigates instruction tuning under multiple layout and stylistic constraints, including tone, structure, and topic coverage. \citet{wan2025cognitive} propose a human-like revision-based framework that reflects realistic writing dynamics under constraint.

Structure-aware evaluation provides critical signals for both training and post-hoc alignment. Long2RAG~\cite{qi2024long2rag} develops key-point recall metrics to assess whether retrieved evidence is adequately incorporated into long-form responses. ExPerT~\cite{salemi2025expert} proposes evaluation strategies that consider layout-specific completeness and explainability. Similarly, \citet{long2025beyond} explore attribute-guided alignment during training to align generation structure with task-specific expectations. \citet{kim2025say} assess verbatim fidelity in long-context models to detect structural drift or misalignment, while \citet{huang2024calibrating} propose calibration methods to correct structural inconsistencies in output generation.

\subsection{Factual Integrity}
Factual consistency is a cornerstone of report generation in Deep Research, where generated outputs must faithfully reflect retrieved evidence. Despite advances in RAG, challenges remain in maintaining fact fidelity, resolving conflicting evidence, and verifying content reliability. To address these issues, recent work explores three key directions: faithful modeling, conflict resolution, and factual evaluation.

Faithful modeling focuses on ensuring that generation aligns with verified and contextually relevant evidence. RAGSynth~\cite{shen2025ragsynth} generates synthetic supervision signals under known factual variations, improving robustness and trustworthiness. BRIDGE~\cite{dai2025after} proposes a verification layer between retrieval and generation to assess factual adequacy. \citet{zhou2023context, shi2024trusting} propose context-aware decoding methods that prioritize high-confidence evidence spans and reduce hallucinated completions, enhancing the factual accuracy of responses.

Conflict resolution is essential when the retrieved sources contain contradicting claims. FaithfulRAG~\cite{zhang2025faithfulrag} introduces fact-level conflict modeling to promote alignment with consistent retrieved facts.  DRAGged~\cite{cattan2025dragged} identifies and mitigates inter-source conflicts using detection and intervention models. Entropy-based decoding strategies~\cite{yuan2024discerning} adaptively adjust to evidence uncertainty, promoting more reliable generation under ambiguous input. \citet{ying2023intuitive} explore LLM behavior under conflicting prompts and improve conflict awareness by behavioral tuning.

Factuality evaluation provides metrics and benchmarks for guiding training and assessing generation reliability. Face4RAG~\cite{xu2024face4rag} evaluates attribution consistency in Chinese RAG systems using fine-grained metrics. SFR-RAG~\cite{nguyen2024sfr} develops contextual attribution indicators to measure alignment at the passage level. \citet{wallat2024correctness} distinguish surface correctness from true source faithfulness. FaithJudge\cite{tamber2025benchmarking} and RAG‑QA Arena~\cite{han2024rag} construct evolving leaderboards and domain-robust benchmarks for long-form factual evaluation. MT‑RAIG~\cite{seo2025mt} extend factuality assessment to structured data contexts, such as multi-table insight generation.

\subsection{Discussion}
Report generation serves as the final step of deep research, transforming fragmented evidence from subproblems into a structured and trustworthy report. To ensure report quality, recent efforts have centered on structure control and factual integrity. Structure control techniques incorporate planning-based generation and constraint-guided decoding, often coupled with structure-aware evaluation and alignment. In parallel, factual integrity is promoted through faithful modeling, which grounds generation in relevant evidence, and conflict reasoning, which enforces consistency across retrieved sources. Together, these approaches aim to produce well-organized and reliable outputs to meet the demands of research-oriented tasks.

Despite this progress, current approaches to report generation in deep research remain in an early stage. Most existing methods target isolated subskills of report generation while lacking joint optimization with upstream components. However, effective report synthesis is inherently coupled with prior planning, question developing, and web exploration stages, making this disconnect a major limitation. Furthermore, current structure control methods often rely on fixed outlines or static planning strategies, lacking the flexibility to adapt to task-specific complexity, evidence distribution, or reasoning flow. On the factuality front, while some methods enhance local grounding, few can model consistency across multi-document and multi-hop contexts, a critical need for scientific research scenarios involving long input contexts and conflicting sources.

\section{Optimization}

\subsection{Workflow}

Deep Research workflows are generally categorized into single-agent and multi-agent systems. Single-agent workflows consolidate all research stages within a unified model, enabling integrated reasoning and end-to-end learning. In contrast, multi-agent workflows decompose the process into specialized modules, promoting parallel execution, modular optimization, and greater flexibility.
% \sout{We categorize Deep Research workflows \yc{workflow or paradigms?} into two main \textcolor{red}{categories}: single-agent and multi-agent systems. Both aim to complete the core stages of Deep Research including planning, question development, web exploration, and report generation, but differ in how these stages are executed and coordinated.}
% \yc{This section is more like another perspective of categorization. Should describe different pipeline design and their corresponding focus?}

\noindent \textit{Single-Agent Systems.} A single LLM agent is responsible for the entire process. Systems such as DeepResearcher~\cite{zheng2025deepresearcher}, WebThinker~\cite{li2025webthinker}, and Search-R1~\cite{jin2025search} typically adopt a monolithic workflow, where the agent sequentially performs task decomposition, query generation, retrieval (via tools or APIs), and final artifact synthesis. All reasoning and decisions are internally managed by the same model, often trained end-to-end using reinforcement learning.

\noindent \textit{Multi-Agent Systems.} In contrast, multi-agent systems assign different agents to specific stages of the pipeline. Planner agents handle task decomposition and subgoal scheduling; query agents focus on generating diversified and contextual queries; retriever agents interact with external tools for evidence gathering; and writer agents perform structured synthesis. This modular division, seen in systems like AgentRxiv~\cite{schmidgall2025agentrxiv}, AI Scientist~\cite{lu2024ai}, and OpenResearcher~\cite{zheng2024openresearcher}, allows each component to be specialized and optimized independently, facilitating parallelism and flexibility in complex research tasks.

\subsection{Parameter Optimization}

Effective Deep Research relies not only on system architecture but also on optimizing agent behavior through training paradigms tailored to complex tasks. Current research focuses on three main optimization approaches: \textit{contrastive learning}, \textit{reinforcement learning}, and \textit{curriculum training}, each addressing different challenges in agent coordination and decision making.
\noindent \textit{Contrastive Learning.} This approach teaches agents to distinguish between effective and ineffective behaviors by contrasting successful and failed trajectories, especially in tool usage such as retrieval, search, and summarization. For example, \textit{Avatar}~\cite{wu2024avatar} trains agents to recognize when and how to invoke external resources, improving precision in multi-step reasoning.
\noindent \textit{Reinforcement Learning.} RL methods fine-tune agents for long-term planning and decision making, using reward signals based on retrieval accuracy and final artifact quality. Systems like \textit{Search-O1}~\cite{li2025search} and \textit{Learning-to-Search}~\cite{chen2025learning} leverage RL to promote effective question formulation and document acquisition in complex research workflows.
\noindent \textit{Curriculum Training.} To handle progressively complex tasks, curriculum training employs staged learning pipelines where agents first master fundamental skills before advancing to full workflow orchestration. Notable examples include \textit{AI Scientist-v2}~\cite{yamada2025ai} and \textit{SimpleDeepSearcher}~\cite{sun2025simpledeepsearcher}, which improve agent robustness in open-ended scenarios.

\noindent \textit{Discussion.} Optimization methods often complement architectural choices: single-agent systems tend to use RL for end-to-end reward alignment, while multi-agent systems benefit more from modular training and agent-level feedback. Some recent frameworks, such as \textit{AgentLab}~\cite{schmidgall2025agent}, begin to explore hybrid optimization combining symbolic planning, search augmentation, and human-in-the-loop feedback.

\begin{table}[ht]
\centering
\caption{Coverage of core Deep Research modules across major benchmarks. P: Planning, QD: Question Developing, WE: Web Exploration, RG:  report generation.}
\label{tab:benchmark}
\begin{tabular}{lccccp{3.5cm}p{4cm}}
\toprule
\textbf{Benchmark} & \textbf{P} & \textbf{QD} & \textbf{WE} & \textbf{RG} & \textbf{Task} & \textbf{Evaluation Metrics} \\
\midrule
\textsc{Mind2Web 2}~\cite{gou2025mind2web}            & \cmark & \cmark & \cmark & \xmark & Web search & Success rate, Partial Completion \\
\textsc{BrowseComp}~\cite{wei2025browsecomp}          & \cmark & \cmark & \cmark & \xmark & Web search & Accuracy, Calibration Error \\
\textsc{WebArena}~\cite{zhou2023webarena}             & \cmark & \cmark & \cmark & \xmark & Web search & Success Rate \\
\textsc{GAIA}~\cite{mialon2023gaia}                   & \cmark & \cmark & \cmark & \xmark & Multi-step assistant tasks & EM \\
\textsc{Humanity’s Last Exam}~\cite{phan2025humanity} & \cmark & \cmark & \cmark & \xmark & Multidomain reasoning & Accuracy, Calibration Error \\
\textsc{BrowseComp-Zh}~\cite{zhou2025browsecomp}      & \cmark & \cmark & \cmark & \xmark & Web search in Chinese & Accuracy, Calibration Error \\
\textsc{MedBrowseComp}~\cite{chen2025medbrowsecomp}   & \cmark & \cmark & \cmark & \xmark & Medical web search & Accuracy \\
\textsc{GPQA}~\cite{rein2024gpqa}                     & \cmark & \cmark & \cmark & \xmark & Graduate-level QA & Accuracy \\
\textsc{InfoDeepSeek}~\cite{xi2025infodeepseek}       & \xmark & \cmark & \cmark & \xmark & Open-domain QA & Accuracy, Information Accuracy \\
\textsc{DeepResearch Bench}~\cite{du2025deepresearch} & \cmark & \cmark & \cmark & \cmark & Research report generation & Pairwise Agreement Rate, Overall Pearson Correlation \\
\textsc{DeepResearchGym}~\cite{coelho2025deepresearchgym} & \cmark & \cmark & \cmark & \cmark & Research task sandbox & KPR, KPC, Precision, Recall, Clarity, Insight \\
\bottomrule
\end{tabular}
\end{table}

\section{Benchmark and Evaluation}

To understand the progress and limitations of Deep Research systems, recent benchmarks have been designed to evaluate the four core technical modules: \textit{Planning}, \textit{Question Developing}, \textit{Web Exploration}, and \textit{Report generation}.  Table~\ref{tab:benchmark} summarizes major benchmarks and the specific modules they cover within the Deep Research workflow. These benchmarks can be broadly categorized into two types based on their task scope: \textbf{search-oriented} and \textbf{research-oriented}.

\textbf{Search-oriented benchmarks} primarily focus on information-seeking tasks involving \textit{Web Exploration} and \textit{Question Developing}, often in interactive or language-specific browsing scenarios. For example, \textsc{Mind2Web 2}~\cite{gou2025mind2web} assesses the ability to reformulate queries and navigate websites dynamically, while \textsc{BrowseComp}~\cite{wei2025browsecomp} and \textsc{BrowseComp-Zh}~\cite{zhou2025browsecomp} evaluate multilingual browsing performance through accuracy and calibration error. \textsc{WebArena}~\cite{zhou2023webarena} extends this with structured tasks that require simple \textit{Planning} components, such as multi-step goal tracking. However, these benchmarks rarely involve final content generation, leaving \textit{ report generation} under-evaluated.

\textbf{Research-oriented benchmarks} go beyond search to evaluate long-range reasoning, synthesis, and structured output production. These tasks require full-pipeline coordination across all four modules. \textsc{DeepResearch Bench}~\cite{du2025deepresearch} and \textsc{DeepResearchGym}~\cite{coelho2025deepresearchgym} are representative, covering everything from \textit{Planning} subtask decomposition, \textit{Question Developing} for subqueries, real-time \textit{Web Exploration}, to final \textit{ report generation} in the form of research-style reports. They provide detailed evaluation metrics like clarity, knowledge precision/recall (KPR/KPC), and agreement scores. \textsc{MedBrowseComp}~\cite{chen2025medbrowsecomp} specializes this pipeline in the biomedical domain, while \textsc{GAIA}~\cite{mialon2023gaia} evaluates agent workflows on general assistant tasks with full modular coverage. Additionally, benchmarks like \textsc{Humanity's Last Exam}~\cite{phan2025humanity} and \textsc{GPQA}~\cite{rein2024gpqa} emphasize \textit{Planning} and \textit{Question Developing} under minimal retrievability, simulating open-book but non-searchable QA scenarios.
\section{Limitations and Future Directions}

Despite impressive progress, Deep Research systems remain in their infancy and face multiple limitations across architecture, reliability, modality, and scalability. We outline several key challenges and future directions below.

\textbf{Multi-Tool Integration.}
Most current systems rely solely on traditional search engines as their primary external tool, which severely restricts their access to diverse and task-specific knowledge sources. However, real-world research often involves querying APIs, parsing structured databases, navigating code repositories, and retrieving information from documents, tables, or charts. Future Deep Research agents must support dynamic orchestration over multiple heterogeneous tools and flexibly decide which to invoke at each reasoning step.

\textbf{Factuality.}
Ensuring factual consistency is a core challenge in Deep Research systems, especially when content is synthesized from multiple sources or spans multi-step reasoning.
% \sout{As Deep Research systems scale to generating multi-paragraph or report-level content, \yc{factuality has no direct relation with report or multi-paragraph content?} ensuring factual consistency becomes increasingly challenging.} 
Agents may inadvertently introduce factual inaccuracies, outdated claims, or unsupported assertions, especially when aggregating content from inconsistent sources. To address this, future systems should incorporate explicit grounding mechanisms, such as source attribution, factuality-aware reward functions, and post-hoc verification modules.

\textbf{Multimodal Reasoning Capabilities.}
Current pipelines are almost exclusively textual, making them unsuitable for domains that require visual or multimodal understanding. Research tasks in science, medicine, and engineering often involve diverse modalities such as images, textual descriptions, and scanned documents. 
Extending Deep Research frameworks to process and reason over multimodal inputs—including images, PDFs, and structured data—remains a largely unexplored but essential direction.

\textbf{Workflow Design and Model Optimization.}
Effective Deep Research requires agents to coordinate complex workflows involving task decomposition, tool usage, and synthesis. 
% However, most current systems rely on fixed, hand-designed prompting or rigid control structures. Future work should explore learnable and adaptive workflows that can evolve with task demands. In parallel, optimizing the underlying LLMs for such workflows remains computationally expensive. Scalable solutions such as parameter-efficient tuning, staged pretraining, and reinforcement learning with human preferences may help mitigate these barriers.
While earlier systems often relied on static prompts and hard-coded workflow, recent works have begun to adopt more adaptive and agentic paradigms, enabling dynamic planning and tool use. However, current methods still lack mechanisms for learning workflows that generalize across tasks and evolve with new objectives. On the optimization side, aligning large language models with such dynamic workflows remains challenging due to high computational costs and sparse reward signals. Future directions may include scalable training strategies such as parameter-efficient tuning, staged or curriculum-based pretraining, and reinforcement learning guided by human or automated feedback.
% \yc{Many existing works are adaptive or agentic.}

\textbf{Personalization}
Personalization aims to align agents with users’ goals and preferences for better performance. Existing methods lack persistent user modeling and dynamic adaptation, often treating personalization as secondary. Future work should develop scalable, privacy-aware user models with continual learning, addressing overfitting and fairness challenges.
% \textbf{Long-Term Complex Task}

% \yc{personalization? long-term complex task?}

In summary, addressing these limitations—through richer tool integration, stronger factual grounding, multimodal expansion, and efficient agent training—will be critical to realizing the full potential of autonomous Deep Research systems.

\section{Conclusion}
Deep research is revolutionizing the search paradigm and has emerged as one of the most promising directions in agent research. In this survey, we provide a systematic overview of the deep research pipeline, which comprises four core stages: planning, question developing, web exploration, and report generation. For each stage, we analyze the key technical challenges and categorize representative methods developed to address them. Furthermore, we summarize recent advances in optimization techniques and benchmarks tailored for deep research. Finally, we discuss open challenges and promising research directions, aiming to chart a roadmap toward building more capable and trustworthy deep research agents.

% The integration of retrieval, reasoning, and generation in autonomous agents marks a new era in language-based research automation. Through this survey, we have provided a structured overview of Deep Research Agents, decomposing their functionality into four foundational modules: \textit{Planning}, \textit{Question Developing}, \textit{Web Exploration}, and \textit{Report generation}. These components collectively enable agents to formulate high-level strategies, decompose complex queries, interact with diverse information sources, and produce high-quality structured outputs.

% We distinguished between \textit{Deep Search}—which enhances retrieval through iterative query refinement and dynamic exploration—and \textit{Deep Research}, which targets end-to-end generation of research artifacts. While significant progress has been made, our analysis of existing benchmarks and systems reveals persistent challenges in tool integration, factual consistency, multimodal reasoning, and workflow optimization.

% As language agents move from task-solving to autonomous knowledge discovery, future efforts must advance beyond isolated improvements and toward holistic, learnable, and reliable research systems. We hope this survey serves as a foundation for future exploration in building next-generation Deep Research Agents.

\bibliographystyle{ACM-Reference-Format}
\bibliography{reference}
% 加方法分类表
% 加framework图
% 加optimization
% section1，2合成一段
\end{document}